# Networks in Cognitive Science

*Andrea Baronchelli[1,*], Ramon Ferrer-i-Cancho[2], Romualdo Pastor-Satorras[3], Nick Chater[4]* and *Morten H. Christiansen[5,6]*

[1] Laboratory for the Modeling of Biological and Socio-technical Systems, Northeastern University, Boston, MA 02115, USA
[2] Complexity & Quantitative Linguistics Lab, TALP Research Center, Departament de Llenguatges i Sistemes Informàtics. Universitat Politècnica de Catalunya, Campus Nord, Edifici Omega, E-08034 Barcelona, Spain
[3] Departament de Física i Enginyeria Nuclear, Universitat Politècnica de Catalunya, Campus Nord B4, E-08034 Barcelona, Spain
[4] Behavioural Science Group, Warwick Business School, University of Warwick, Coventry, CV4 7AL, UK
[5] Department of Psychology, Cornell University, Uris Hall, Ithaca, NY 14853, USA
[6] Santa Fe Institute, 1399 Hyde Park Road, Santa Fe, NM 87501, USA
[*] Corresponding author: email a.baronchelli.work@gmail.com

**Abstract**

Networks of interconnected nodes have long played a key role in cognitive science, from artificial neural networks to spreading activation models of semantic memory. Recently, however, a new Network Science has been developed, providing insights into the emergence of global, system-scale properties in contexts as diverse as the Internet, metabolic reactions or collaborations among scientists. Today, the inclusion of network theory into cognitive sciences, and the expansion of complex systems science, promises to significantly change the way in which the organization and dynamics of cognitive and behavioral processes are understood. In this paper, we review recent contributions of network theory at different levels and domains within the cognitive sciences.

Humans have more than $10^{10}$ neurons and between $10^{14}$ and $10^{15}$ synapses in their nervous system [1]. Together, neurons and synapses form neural networks, organized into structural and functional sub-networks at many scales [2]. However, understanding the collective behavior of neural networks starting from the knowledge of their constituents is infeasible. This is a common feature of all complex systems, summarized in the famous motto "more is different" [3]. The study of complexity has yielded important insights into the behavior of complex systems over the past decades, but most of the toy models that proliferated under its umbrella have failed to find practical applications [4]. However, in the last decade or so a revolution has taken place. An unprecedented amount of data, available thanks to technological advances, including the Internet and the Web, has transformed the field. The data-driven modeling of complex systems has led to what is now known as Network Science [5].

Network Science has managed to provide a unifying framework to put different systems under the same conceptual lens [5], with important practical consequences [6]. The resulting formal approach has uncovered widespread properties of complex networks and led to new experiments [4] [7] [8]. The potential impact on cognitive science is considerable. The newly available concepts and tools already provided insights into the collective behavior of neurons [9], but they have also inspired new empirical work, designed, for example, to identify large-scale functional networks [10] [11]. Moreover, very different systems such as semantic networks [12], language networks [13] or social networks [14, 15] can now be investigated quantitatively, using the unified framework of Network Science.

These developments suggest that the concepts and tools from Network Science will become increasingly relevant to the study of cognition. Here, we review recent results showing how a network approach can provide insights into cognitive science, introduce Network Science to the interested cognitive scientist without prior experience of the subject, and give pointers to further readings. After a gentle overview of complex networks, we survey existing work in three subsections, concerning the neural, cognitive, and social levels of analysis. A final section considers dynamical processes taking place *upon* networks, which is likely to be an important topic for cognitive science in the future.

## I. Introduction to Network Science

The study of networks (or graphs) is a classical topic in mathematics, whose history began in the 17th century [16]. In formal terms, networks are objects composed of a set of points, called vertices or nodes, joined in pairs by lines, termed edges (see Fig. 1 for basic network definitions). They provide a simple and powerful representation of complex systems consisting of interacting units, with nodes representing the units, and edges denoting pairwise interactions between units. Mathematical graph theory [17], based mainly on the rigorous demonstration of the topological properties of particular graphs, or in general extremal properties, has been dramatically expanded by the recent availability of large digital databases, which have allowed exploration of the properties of very large real networks. This work, mainly conducted within the statistical physics community, has led to the discovery that many natural and artificial systems can be usefully described in terms of networks [8]. The new Network Science has been successfully applied in fields ranging from computer science and biology to social sciences and finance, describing systems as diverse as the World-Wide Web, patterns of social interaction and, collaboration, ecosystems, and metabolic processes (see [7] for a review of empirical results).

Interest in real complex networks has been boosted by three empirical observations. The first is the so-called *small-world* effect, first observed experimentally by the social psychologist Stanley Milgram [18], which implies that there is a surprisingly small shortest path length, measured in traversed connections in direct paths, between any two vertices in most natural networks. In Milgram's experiment, a set of randomly chosen people in Omaha, Nebraska, were asked to navigate their network of social acquaintances in order to find a designated target, a person living in Boston, Massachusetts. The navigation should be performed by sending a letter to someone the recipients knew on a first-name basis, which they thought should be closer to the target, and asking them to do the same until the target was reached. The average number of people that the letters passed through before reaching the target led to the popular aphorism "six degrees of separation". While the number six is not universal, the average distance between pairs of vertices in real networks is typically very small in relation to network size.

The second observation concerns the high *transitivity* of many real networks. The concept of transitivity is borrowed from usage in the social sciences [19], and refers to the fact that, for example, two friends of any given individual are themselves also likely to be friends. Transitivity can be quantitatively measured by means of the clustering coefficient [20], which takes large values in almost all real networks.

Thirdly, the connectivity structure of many real systems is strongly heterogeneous, with a skewed distribution in the number of edges attached to each vertex (the so-called degree distribution) (Fig. 2 and 3). This kind of networks has been dubbed *scale-free* [21]. The scale-free hallmark underlies many of the most surprising properties of complex networks, such as their extreme resilience to random deletion of vertices, coupled with extreme sensitivity to the targeted deletion of the most connected vertices [22]; and it strongly impacts processes such as the propagation of disease [23].

## II. Applications of network theory in Cognitive Science

***The brain and neural networks.*** The network framework provides a natural way to describe neural organization [24]. Indeed, cognition emerges from the activity of neural networks that carry information from one cell assembly or brain region to another (see Box II). The advent of Network Science suggests modifying the traditional "computer metaphor" for the brain [25] to an "Internet metaphor", where the neocortex takes on the task of "packet switching' [26].

More broadly, network theory allows the shift from a reductionist to a "complex system" view of brain organization [2, 9, 10, 27]. In this framework, optimal brain functioning requires a balance between local processing and global integration [28, 29]. In particular, clustering facilitates local processing, while a short path length (a low degree of separation) across the neural network is required for global integration of information among brain regions. Indeed, these two factors may shape neural network structure and performance [30, 31].

The map of brain connectivity, the so-called connectome (see Glossary), and its network properties are crucial for understanding the link between brain and mind [29]. The connectome is characterized by short path lengths (a small-word topology), high clustering, and *assortativity,* the tendency of hubs to be connected to hubs, forming a so-called 'rich club', and an overlapping community structure [32-35]. The latter observation challenges earlier reductionistic views of the brain as a highly modular structure (e.g., [36]).

Alterations of fundamental network properties are often associated with pathologies [28, 37, 38] [39]. For instance, smaller clustering, larger path length and greater modularity are found in autistic spectrum disorder [38]. Similarly, the multimodal cortical network has a shorter path length and a trend to increased assortativity in schizophrenics [37]. It is not clear if Alzheimer's disease has a unique signature at the brain network level, but in different studies path lengths and clustering have been found to be altered, both above or below controls [28].

Intriguingly, Network Science may provide the tools to describe different kinds of brain networks in a coherent fashion, and compare their properties even across different scales. Particularly remarkable is the identification of *large-scale* brain networks, defined according to structural connectivity or functional interdependence [27] [10]. The network approach has also been a driving force in the analysis of functional networks in neuroimaging data [2]. For example, fMRI techniques, an indirect measure of local neuronal activity [40], have shown dynamic reconfiguration of the modular organization of large-scale functional network during learning [41]. Moreover, various pathologies have been related to alterations of the properties of large-scale networks [10]. Different neurodegenerative diseases have been connected with the degradation of different large-scale functional networks [42], while age-related changes in face perception have been linked to the degeneration of long range axonal fibers [43].

***Cognitive processes.*** At the level of cognition (i.e., the information-processing operations in the brain), a wide range of networks has been considered [44-50]. One of the most studied examples are networks of free word associations, which are in general weighted and directed, with weights reflecting the frequency of a given association [12, 51]. Short path lengths, high clustering and assortativity have been reported across datasets [44, 52]. High clustering and short path lengths have been attributed to a network dynamics combining 'duplication' and 'rewiring' (Box I).

A key theoretical question is whether the properties of networks at the level of information processing are inherited from the brain network substrate or instead arise from independent converging processes [10]. Cognitive impairment was found to be associated with a drop of path lengths and a rise of clustering in word fluency networks in Alzheimer patients [46], whereas the opposite trend (increased path lengths and decreased clustering) was found in associative networks of late talkers [53]. Understanding the relationship, if any, between these alterations at the cognitive and neural levels is a challenge for future research.

Network Science has also shown how to single out the most important elements of a complex system. The simplest approach focuses on the concept of "degree:" "hubs" are highly connected nodes whose removal causes greater impact than low degree nodes [22]. The internal organization of cognitive networks has been analyzed also at a larger scale, identifying the network's "core" [54-56] and dividing ensembles of nodes into "communities" that map into semantic [57, 58] or syntactic [45] categories. It has been hypothesized that the lexicon may contain a basic vocabulary from which the meaning of the remaining words can be covered via circumlocution [59] [60]. This hypothesis has been supported by the analysis of language networks [13] of word co-occurrence in many languages [61, 62] and web search queries [63], where the degree distribution shows a power law with two regimes, one containing essential vocabulary and the other containing specialized terms. The two regimes may emerge naturally from a type of preferential attachment dynamics [55] (see Box I). Similarly, a network analysis of cross-referencing between dictionary entries has shown that dictionaries have a so-called grounding kernel, a subset of a dictionary consisting of about 10% of words (typically with a concrete meaning and acquired early), from which other words can be defined [54].

As far as semantics is concerned [12], in word association networks, names of musical instruments or color terms form strongly interconnected subsets of words, i.e., communities of nodes [57, 58]. Similarly, parts of speech (e.g., verbs and nouns) cluster together in a syntactic dependency network [45]. This organization may help explain why brain damage can affect particular semantic fields [64] or specific parts-of-speech [65].

Network theory offers many new perspectives for understanding cognitive complexity. The ease with which a word is recognized depends on its degree or clustering coefficient [66-68]. Network theory has also helped to quantify the cognitive complexity of navigating labyrinths, whose structure, including the distance between relevant points, can be coded as a weighted network, distinguishing purely aesthetic labyrinths from those that were designed to have a complex solution [69]. The time needed to find the way out of a labyrinth is strongly correlated with that needed by a random walker (Box III) to reach the exit (absorption time), which is in turn strongly correlated with the various network metrics including vertex strength and betweenness [69]. An interesting possible research direction is to investigate whether similar analysis applies to search problems in more abstract cognitive contexts, such as problem solving or reasoning.

The study of sequential processing has also been impacted by Network Science. For example, the length of a dependency between two elements of a sequence provides a measure of the cognitive cost of that relationship [70]. Thus, the mean of such lengths may measure the cognitive cost of process a sequence, such as a sentence [71, 72]. The minimum linear arrangement problem is to determine the ordering of elements of the sequence that minimizes such sum of lengths, given a network defining the dependencies between elements (Box VI, Fig 4 [71, 73, 74]. The rather low frequency of dependency crossings in natural language (Fig. 4 (c) versus (d)) and related properties could be a side effect of dependency length minimization [74-76] suggesting that crossings and dependency lengths cannot be treated as independent properties, as it is customary in cognitive sciences [71, 77]. These findings suggest that a universal grammar is not needed to explain the origins of some important properties of syntactic dependencies structures: the limited capacity of the human brain may severely constrain the space of possible grammars. The network approach additionally allows for a reappraisal of existing empirical evidence. For example, the second moment of the degree distribution, $<k^2>$, is positively correlated to the minimum sum of dependency lengths (Box VI), and therefore sufficiently long sentences cannot have hubs [78]. While the minimum linear arrangement problem has so far been investigated mostly in language, it applies whenever a dependency structure over elements of a sequence is defined by a network. A promising avenue for future research is to extend network analysis to sequences of non-linguistic behavior, such as music, dance and action sequencing.

Various studies address the origin of the properties of cognitive networks (Box I). For example, the double power law degree distribution observed in word co-occurrence networks, with two different exponents, has been attributed to a dynamics combining the growth and preferential attachment rules, where a pair of disconnected nodes becomes connected with probability proportional to the product of their degrees [55]. The model is only a starting point, as it fails to reproduce other important properties of the real networks, e.g. the distribution of eigenvalues of the corresponding adjacency matrix [61]. A different model, not based on preferential attachment, and mirroring a previous model of protein interaction networks [79], introduced the concepts of growth via node duplication and link rewiring to cognitive science, to provide a unified explanation of the power-law distribution, the short path length, and the high clustering of semantic networks [44]. However, a simple network growth dynamics is not necessarily the best mechanism. In a network of Wikipedia pages, the distribution of connected component sizes at the percolation threshold was found to be inconsistent with a randomly growing network [80]. In phonological similarity networks, five key properties—the largest connected component including about 50% of all vertices, small path lengths, high clustering, exponential degree distribution and assortativity [81]—may arise from a network of predefined vertices and connections defined simply by overlap between properties of the node, rather than a growth model [82]. Overall, the debate over the different origins of cognitive networks highlights the importance of defining suitable model selection methods (see Section IV).

The network approach suggests potentially revolutionary insights also into the fast or even abrupt emergence of new cognitive functions during development as well as the degradation of those functions with aging or neurodegenerative illness. Such abrupt changes can arise from smooth change, if the system crosses a percolation threshold, i.e., a crucial point where the network becomes suddenly connected (e.g., during development) or disconnected (during aging or illness). The existence of such a point has been demonstrated in a semantic network extracted by Wikipedia evolving by the addition of new pages [80]. Furthermore, the concept of percolation has inspired a recent explanation of hyperpriming and related phenomena exhibited by Alzheimer's disease patients in a theoretical model that qualitatively captures aspects of the experimental data [83].

***Social networks and cognition.*** Network Science has been fruitfully applied to the investigation of networks of interactions between people, highlighting the interplay between individual cognition and social structure. For example, collaboration networks, both in scientific publications [84] and in Wikipedia [85], where a link is established between two authors if they have collaborated on at least one paper or page, provide insights into the large-scale patterns of cooperation among individuals, and show a pronounced small-world property and high clustering. [84]. Similarly, a 'rich get richer' phenomenon turns out to drive the dynamics of citation networks, both between papers and authors [21]. Scientific authors tend to cite already highly cited papers, leaving importance or quality in second place [86]. Moreover, pioneer authors benefit from a "first-mover advantage'" according to which the first paper in a particular topic often ends up collecting more citations than the best one [87]. The same approach has also allowed identifying the mechanisms that govern the emergence of (unfounded) authority among scientists, and their consequences [88].

One recent focus of research has been the large-scale validation of the so-called Dunbar number. Dunbar compared typical group size and neocortical volume in a wide range of primate species [89], concluding that biological and cognitive constraints would limit the immediate social network of humans to a size of 100-200 individuals [90]. Analyzing a network of Twitter conversations involving 1.7 million individuals, it has been possible to confirm that users can maintain a limited number of stable relationships, and that this number agrees well with Dunbar's predictions [91].

Social networks play a fundamental role also in collective problem-solving tasks [92]. For example the speed of discovery and convergence on an optimal solution is strongly affected by the underlying topology of the group in a way that depends on the problem at hand [14] [93]. More spatially based cliques seem to be advantageous for problems that benefit from broad exploration of the problem space whereas long distance connections enhance the results in problems that require less exploration [14], even though recent experiments suggest that long distance connections might always be advantageous [93]. Similarly, the amount of accessible information impacts problem solving in different ways on different social network structures, more information having opposite effects on different topologies [94].

Human behavior in social interactions has been revealed through the empirical analysis of phone calls [95, 96] and face-to-face interaction networks [97, 98]. This research has clarified the relationship between the number and the durations of individual interactions, or, put in network terms, between the degree and the strength of the nodes. Surprisingly, it turns out that this relation differs in phone vs face-to-face interactions: the more phone calls an individual makes, the less time per call he or she will allot [99], but for face-to-face interactions, popular individuals are "super-connectors," with not only more, but also longer, contacts [98]. Other insights into the effect of social networks have been obtained through controlled experiments on the spread of a health behavior through artificially structured online communities [100]. Behavior spreads faster across clustered-lattice networks than across corresponding random networks. The impacts of network structure in understanding how societies solve problems and passing information may have strong parallels with how the "society of mind" [101] within a single individual is implemented in information processing mechanisms and neural structure.

**III. Simple dynamics on networks**.

So far we have considered the structure of networks and the dynamical principles of growth or deletion (re)shaping these structures. Recently, however, new approaches have adopted a different perspective [102]: The neural, cognitive or social process is modeled as a *dynamic* process taking place *upon* a network. Researchers can then ask how the network structure affects the dynamics.

An illustrative example concerns the interactions among neural or cortical neurons, which often yield network level synchrony [103-105]. Various studies reveal that abnormal synchrony in the cortex is observed in different pathologies, ranging from Parkinson's disease (excessive synchrony) [106] to autism (weak synchrony) [107] [108]. Neural avalanches constitute another important process occurring on brain networks [109]. The size distribution of these bursts of activity approximate a power law, often a signature of complex systems [110]. The Kinouchi-Copelli (KC) model suggested that the neuronal dynamic range is optimized by a specific network topology tuned to signal propagation among interacting excitable neurons, and which leads to  neural synchronization as a side-effect [111]. Remarkably, the predictions of this model have been confirmed empirically in cultures of cortex neurons where excitatory and inhibitory interactions were tuned pharmacologically [112]. Similar phenomena have been identified in connection to maximal synchronizability [104], information transmission [109, 113] and information capacity  [113] in cortical networks.

In the same way, it is interesting to speculate that some aspects of memory, thought and language may be usefully modeled as navigation (i.e. the process of finding the way to a target node efficiently [114, 115] [48]) or exploration (i.e., navigation without a target) on network representations of knowledge by means of various strategies, such as simple random walks [58, 116] or refined versions combining local exploration and 'switching' [117]. Statistical regularities such as Zipf's law can arise even from a random walk through a network where vertices are words [116]. Semantic categories and semantic similiarity between words can then emerge from properties of random walks on a word association network [58]. Improved navigation strategies (random walks with memory) help to build efficient maps of the semantic space [118]. Furthermore, people apparently use nodes with high closeness centrality to navigate from one node to another in an experiment on navigating an artificial network [48] [119]. These nodes are reminiscent of the landmarks used to navigate in the physical environment [120].

Network analysis casts light on the so-called "function" words [121] (*in*, *the*, *over*, *and*, *of*, etc). These are hubs of the semantic network and they are indeed 'authorities' according to PageRank, a sophisticated technique used by Google to determine the importance of a vertex (e.g., a word) from its degree and the importance of its neighbors [47]. Such hubs provide efficient methods for the exploration of semantic networks [117]. Moreover, the ease with which a word is recognized depends on its degree [66] and its clustering coefficient [67] [68]. PageRank turns out to be a better predictor of the fluency with which a word is generated by experimental participants than the frequency or the degree of a word [119].

Another example is found in the collective dynamics of social annotation [122], occurring on websites (such as *Bibsonomy*) that allow users to tag resources, i.e., to associate keywords to photos, links, etc. First, a co-occurrence graph is obtained by establishing a link between two tags if they appear together in at least one post. The study of the network's evolution generates interesting observations, such as yet another power law, Heaps' law, which relates the number of word types ("the observed vocabulary size") and word tokens in a language corpus [123]. In addition, the mental space of the user is represented in terms of a synthetic semantic network, and a single synthetic post is then generated by finite random walk (see Box III) exploring this graph. Many synthetic random-walk-generated posts are then created, and an artificial co-occurrence network is built. Different synthetic mental spaces are then tested. The artificial co-occurrence network turns out to reproduce many of the features of the real graph, if the synthetic semantic graph has the small-world property and finite connectivity [122].

In the study of language dynamics and evolution, social networks describing the interactions between individuals have been central [15, 124]. The role of the topology of such networks has been studied extensively for the Naming Game [125, 126], a simple model of the emergence of shared linguistic conventions in a population of individuals. When the social network is fully connected, the individuals reach a consensus rapidly, but the possibility of interacting with anybody else requires a large individual memory to take into account the conventions used by different people [126]. When the population is arranged on a lattice, on the other hand, individuals are forced to interact repeatedly with their neighbors [127], so that while local agreement emerges rapidly with the agents using a very little memory, global convergence is reached slowly through the competition of the different locally agreeing groups (local clusters). Small-world networks, on the other hand, turn out to be optimal in the sense that finite connectivity allows the individuals to use a finite amount of memory, as in lattices, while the small-world property prevents the formation of local clusters [15, 128]. Similar analyses have been performed for the case of competition not between specific linguistic conventions, but between entire languages [129, 130] [131]. Overall, these studies highlight the importance of the properties of social networks for the emergence and maintainance of complex cognition, language and culture. The study of dynamics on networks is also likely clarify the relevance of properties of network structure, such as path lengths and clustering, for cognitive processes and their pathologies. A take home message from this research is that network theory challenges radically the view that the unique requirement for complex cognition and its multiple manifestations is the human brain. Instead, the key for the emergence and maintenance of such skills might be the properties of the network defining how the individuals interact.

## IV. Methodological issues for future research

In spite of the enormous potential of network theory for the cognitive and brain sciences, important methodological challenges remain, regarding network construction, analysis and modeling.

***Challenges for network construction.*** A basic challenge for the analysis of co-occurrence networks is determining if two vertices have co-occurred above chance—i.e., inferring whether an edge should be drawn between them (an issue that arises in the parallel literature on probabilistic graphical models [160,161]). In networks of co-occurrence typically no statistical filter is used [132] or the filter is not well-defined [62]. For this reason, proper statistical filters (e.g., [133]) or more precise ways of linking vertices have been considered, e.g. syntactic dependency instead of word-occurrence [134, 135]. In general, however, defining an appropriate null hypothesis for the existence of an edge is crucial. In case of networks induced from co-occurrences of elements in a sequence [132, 136], this would distinguish between significant above-random properties, identified by the ensemble of permutations of the original sequence (e.g., the permutations of a text) and non-significant findings. Even the latter are important, however, as they may suggest that some features of the network could be a side-effect purely of the frequency with which the elements occur. The issue is a general one: in brain network research it arises when determining if the activity of two brain regions is really correlated [137], while in collaboration networks, connecting two scientists because they have "co-occurred" in the coauthor list of an article does not imply that they have actually collaborated [84]. This variety of applications highlights the value of Network Science in offering a unified framework to the various areas of Cognitive Science.

***Challenges for network analysis.*** The most commonly used null hypothesis for the statistical properties of a network is the Erdös-Rényi (or binomial) network (Box I, Fig. 2 and 3). A better null hypothesis is a network that preserves the original degree sequence but in which edges are randomized [8], which in general clarifies the role of the degree distribution and how it could be responsible for the properties of the observed network. For instance, a power-law distribution of degrees may lead to an apparently large clustering coefficient, in networks of not too large a size [8]. Other properties, however, can depend on further details apart from the degree distribution. For example, apparently harmless manipulations such as banning loops (edges from a node to itself) and multiple edges (more than two edges joining a pair of nodes) can lead to degree correlations and disassortative behavior in power-law degree distributions (see Glossary) [138].

Another challenging problem concerns the degree distribution, which is often assumed to be a power-law (Box I). First of all, where a power-law is certain (Fig. 3), direct regression methods to determine the degree exponent are potentially biased and an estimation by maximum likelihood is more convenient [139]. However, various distributions, not only the power law, are able to mimic an approximate straight line in double logarithmic scale [140, 141], and a power law degree distribution has been found not to be sufficiently supported in biological networks, contrary to previous beliefs [142, 143]. In general, the analysis of the degree distribution would require the use of standard model selection techniques from an ensemble of candidate distributions [144]. Equivalent evaluations of a power-law in cognitive networks are not available as far as we know.

***Challenges for dynamical models.*** A big challenge for understanding the dynamical processes underlying brain and cognitive networks is determining which underlying network model is most appropriate. Since many network models can account for a power law distribution (Box I), other network features must be introduced in the evaluation of the most likely model. However, perhaps the most valuable information is how the network has evolved to reach a certain configuration. Different dynamical rules may lead to the same end-product and it is possible to use sophisticated techniques to assess the importance of different evolutionary mechanisms [145-147]. These methods could help clarify the debate on the actual dynamical principles guiding the evolution of semantic networks, e.g. preferential attachment and its variants, in normal and late talkers [53, 148]. Incorporating the statistical methods mentioned above is vital to harness the power of network science to reveal the dynamical principles by which brain is structured and by which brain functions emerge, develop and decay.

## V Conclusions and Outlook

Our survey of the vast literature of network theory for brain and cognitive sciences has necessarily been selective, but it allows us to draw several encouraging conclusions. Network Science offers concepts for a new understanding of traditional terms in cognitive science (Table I) and illuminates a wide range of phenomena, such as the organization of pathological brains or cognition (e.g., [38]), the development of vocabulary in children (e.g., [47, 148]) or language competition, (e.g. [131]) under the same theoretical umbrella. Many new questions arise concerning how far network properties at the neural level translate into network properties at higher levels and vice versa (see Outstanding Questions). Network theory also may help bridge the gap between the brain and the mind, shedding new light on how knowledge is stored and exploited, as well as reduce the gulf that separates the study of individual and collective behavior. Moreover, understanding the origin of the observed properties of networks through the tools of Network Science may help unify research on the development of cognition during childhood with the study of processing in the adult state and its decay during aging or illness. Network science is a young discipline (Box V), but it promises to be a valuable integrative framework for understanding and relating the analysis of mind and behavior at a wide range of scales, from brain processes, to patterns of social and cultural interaction. Overall, network theory can help cognitive science become more internally coherent and more interconnected with the many other fields where network theory has proved fruitful.

## Acknowledgements

We are grateful to B. Elvevåg for making us aware of relevant work and helpful discussions. R. Ferrer-i-Cancho was supported by the grant Iniciació i reincorporació a la recerca from the Universitat Politècnica de Catalunya and the grants BASMATI (TIN2011-27479-C04-03) and OpenMT-2 (TIN2009-14675-C03) from the Spanish Ministry of Science and Innovation (RFC). R. Pastor-Satorras acknowledges financial support from the Spanish MICINN, under project FIS2010-21781-C02-01 and additional support through ICREA Academia, funded by the Generalitat de Catalunya. Nick Chater is supported by ERC Advanced Grant 295917-RATIONALITY.

**FIGURES**

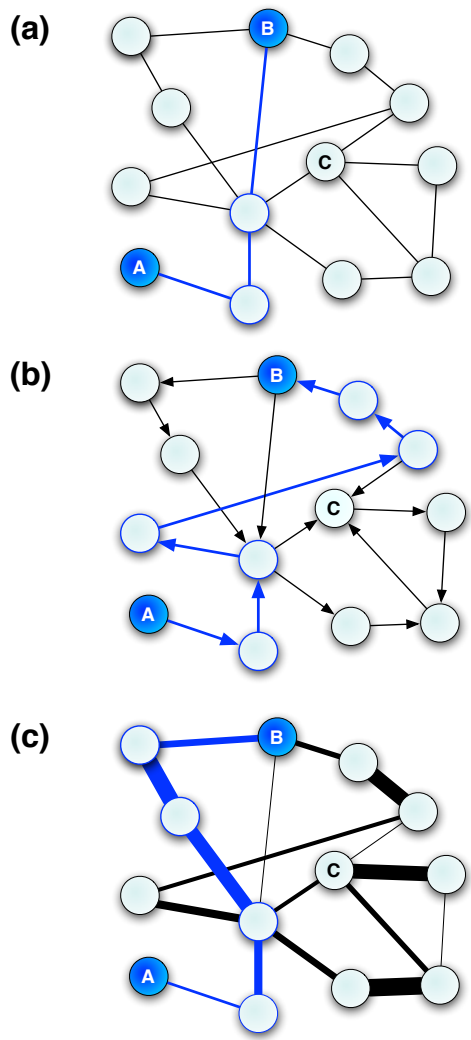

**(a)**

**(b)**

**(c)**

FIGURE 1: *Basic network properties*. (a). Circles represent vertices, while solid lines connecting pairs of vertices correspond to edges. The degree *k* of a vertex is given by number of its neighbors, i.e., the number of other vertices to which it is connected by edges. For example, node *C* in the Figure has degree *k=4*. The distance (shortest path length) $\ell$ between two nodes is given by the minimum number of edges that connect them in a continuous path. For example nodes *A* and *B* are at distance $\ell = 3$ . (b). In a directed network, vertices are unidirectional, indicating that the flow of information can only proceed in one direction between adjacent nodes. The distance between nodes *A* and *B* is now $\ell = 6$. As for node *C*, it has in-degree $k_{in}=3$ and out-degree $k_{out}=1$, meaning that it can receive information from three nodes and pass it to just one neighbor. (c). In a weighted network, links have different capacities, or weights, indicating the amount of information they can carry. Many definitions of distance can be adopted. In the figure, the path between nodes *A* and *B* highlighted in blue is obtained by following the maximum weight link at each step. Beyond its degree, a node is characterized also by its strength, i.e. the sum of the weights of the links that connect it to the rest of the network.

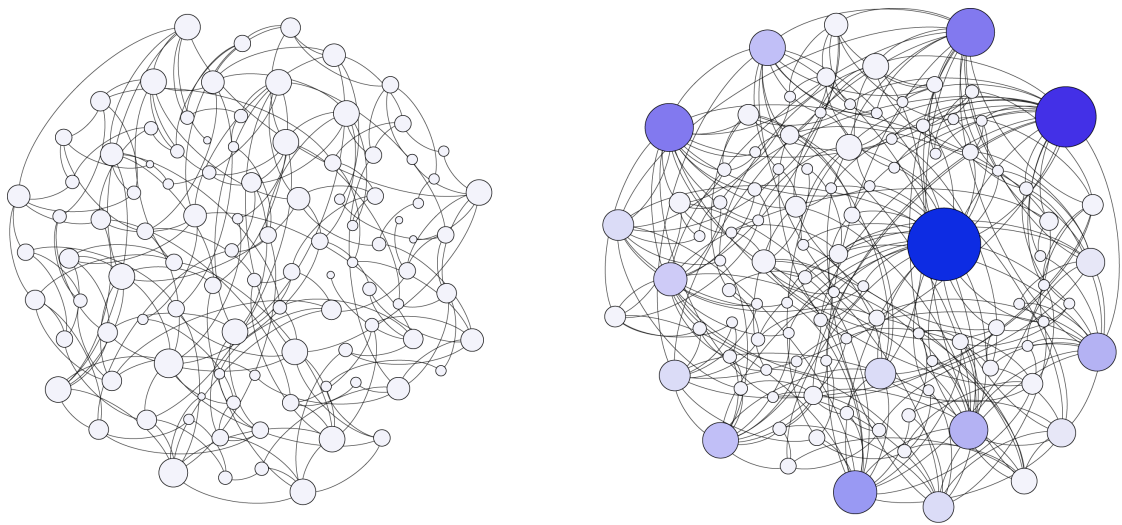

FIGURE 2: *Graphical representation of homogeneous and scale-free networks.* In homogeneous networks (left) nodes have similar topological properties, which are well captured by their average values. In heterogeneous networks (right), on the other hand, very different nodes coexist, including some so-called hubs, i.e., extremely well connected nodes. In both figures, the degree of each node is visually stressed by color and size. The left panel depicts an Erdös-Rényi random graph, the right one a Barabasi-Albert graph, both containing *N=100* nodes and the same average degree *<k>=2.5.*

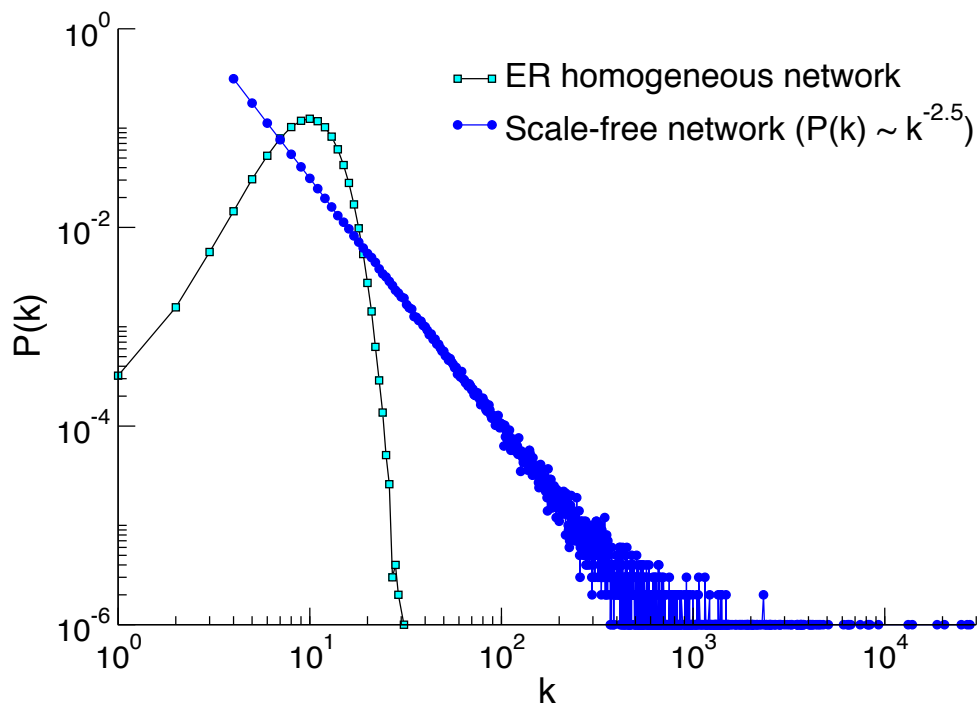

FIGURE 3: *Degree distribution.* The degree distribution of a network, *P(k)*, tells us the probability that a randomly chosen node will have degree *k*. In the figure, the degree distribution of an Erdös-Rényi (ER) graph is plotted next to the one of a scale-free network (with *P(k)~k⁻²·⁵*). It is clear that, while in the ER graph the probability of observing a node with degree *k>30* is practically zero, in the scale-free graph there is a finite, if small, probability to observe hubs connected to thousands, or even tenth of thousands, nodes. Both graphs have the same size *N=10⁶* nodes, and the same average degree *<k>=10.5*

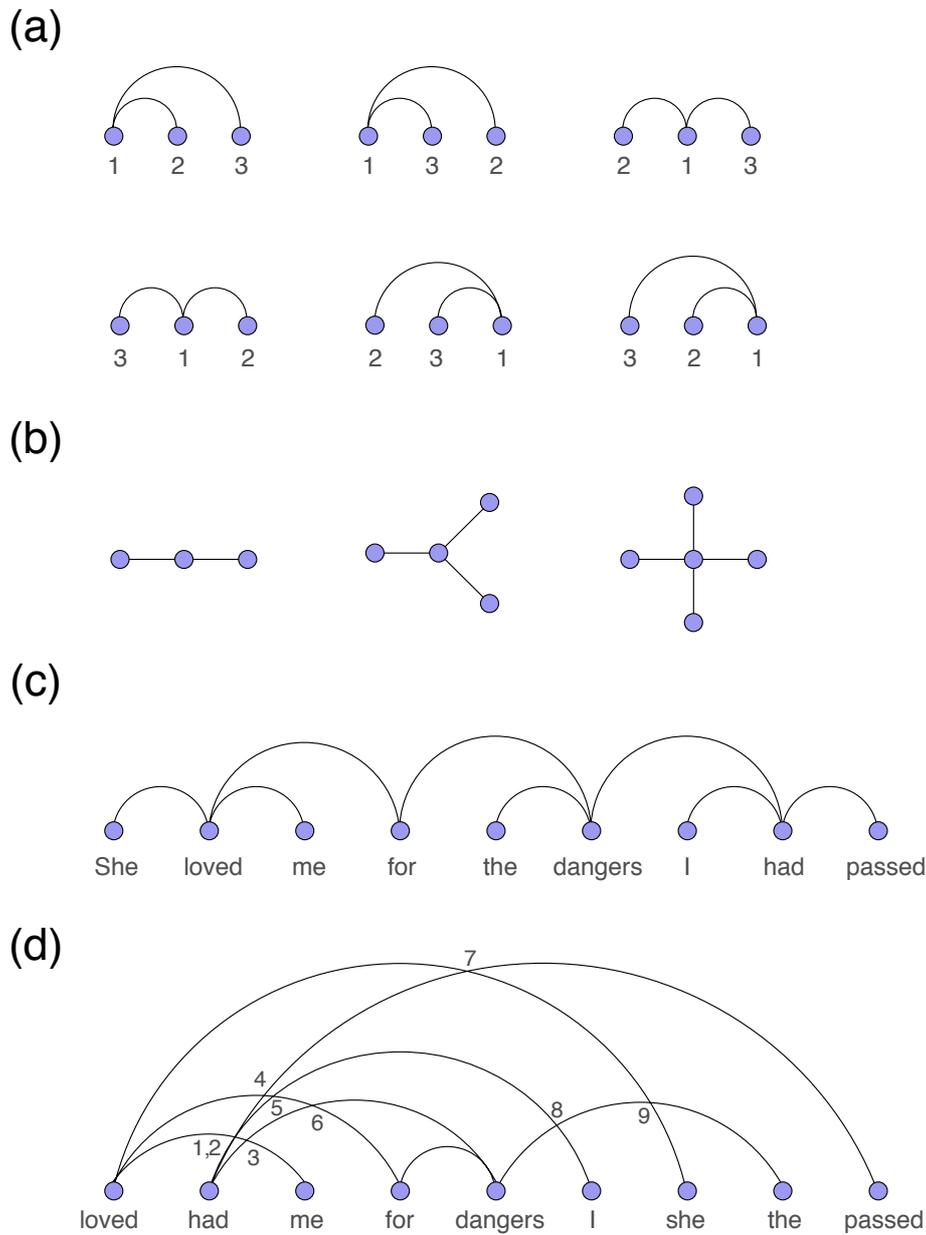

FIGURE 4. (a). The six possible linear arrangements of the three vertices of a tree. (b). Star trees of three, four and five vertices. (c). The syntactic dependency tree of an English sentence (borrowed from [149]). Vertices are words and edges indicate syntactic dependencies between words. (d). A random linear arrangement of the sentence (c) with nine edge crossings indicated with numbers from 1 to 9 (adapted from [74]). Two edges cross if they do not share vertices and one of the vertices making one of the edges is placed between the pair of vertices making the other edge. For instance, the 7-th crossing is formed by the edge between "loved" and "she" and the edge between "had" and "passed".

**GLOSSARY**

**Assortativity:** preference for the nodes to attach to other vertices that are similar in some way. Assortative mixing by degree, for example, describes the case in which nodes that have similar degree tend to form connections preferentially among themselves. For example, in most social networks high degree nodes are connected to high degree nodes, and at the same time also poorly connected vertices tend to link each other. By contrast, disassortativity describes the opposite tendency (e.g., high degree nodes tend to be connected to low degree nodes in technological networks).

**Betweenness of a node:** the number of shortest paths between pairs of vertices passing through a given vertex.

**Clique:** subset of a network where every pair of nodes is connected.

**Clustering coefficient:** $c_i$ of vertex $i$ is defined as the ratio between $e_i$, the actual number of edges between its nearest neighbors and the maximum possible number $k_i(k_i - 1)/2$, i.e.

$$c_i = \frac{2e_i}{k_i(k_i - 1)}.$$

The clustering coefficient quantifies the *transitivity* of a network, measuring the probability that two vertices with a common neighbor are also neighbors of each other. The average clustering coefficient $\langle c \rangle$ is the average value of $c_i$ over all vertices in a network, i.e.

$$\langle c \rangle = \frac{1}{N} \sum_i c_i.$$

In real networks, $\langle c \rangle$ takes usually values of order unity, in stark contrast with the clustering inversely proportional to network size that is expected for a random network (Box I).

**Centrality:** the centrality of a node measures its relative importance inside the network, e.g. in terms of degree, betweenness or distance. In the latter case, the so-called closeness centrality is defined as the inverse of the sum of the shortest path lengths from the considered vertex to all other vertices in the network.

**Community:** although the precise definition of community is still an open question, a minimal and generally accepted description is that the subset of the nodes in a community is more tightly connected to one another than to the rest of the network.

**Connected component:** maximal subset of vertices in a network such that there is a path joining any pair of vertices in it.

**Connectome:** the detailed "wiring diagram" of the neurons and synapses in the brain.

**Co-occurrence network**: a network with nodes representing the elements present in a given context, e.g., words in a text, and edges representing the "co-occurrence" in the same context, according to some criterion; for example, in the case of words in a text, a simple criterion is that they appear in the text one next to the other.

**Core of a network:** a powerful subset of the network because of the high frequency of occurrence of its nodes [55], their importance for the existence of remainder of nodes [54], or the fact that is both densely connected and central (in terms of graph distance) [56].

**Degree:** of a vertex, $k_i$, is defined as the number of other vertices to which vertex $i$ is connected (or "number of neighbors" [66]).

**Degree distribution:** the probability $P(k)$ that a randomly chosen vertex has degree $k$ for every possible $\underline{k}$. For large networks, the degree distribution represents a convenient statistical characterization of a network's topology.

**Diameter:** Longest of the shortest paths between any pair of vertices in a network.

**Directed network**: a network in which each link has an associated direction of flow.

**Hubs:** are the vertices in a network with the largest degree (number of connections).

**Network or graph:** collection of points, called vertices (or nodes), joined by lines, referred as edges (or links). Vertices represent the elementary components of a system, while edges stand for the interactions or connections between pairs of components.

**PageRank:** network analysis algorithm that assigns a numerical weight to each edge of a directed network, aimed at measuring its relative importance. The algorithm is used by the Google search engine to rank Word-Wide Web search results.

**Percolation threshold**: percolation theory describes the behavior of connected clusters in a graph. A network is said to percolate when its largest connected component contains a finite fraction of the nodes that form whole network. Percolation depends in general on some topological quantity (e.g. the average degree in the Erdös-Rényi random graph). The percolation threshold is the value of this quantity above which the network percolates.

**Rich-club phenomenon:** property observed in many real networks in which the hubs have a strong tendency to be connected to each other, rather than with vertices of small degree.

**Scale-free networks:** networks with a broad, heavy-tailed degree distribution, which can often be approximated by a power-law, $P(k) \sim k^{-\gamma}$, where $\gamma$ is a characteristic exponent usually in the range between 2 and 3. This heavy-tailed power-law form underlies many of the surprising features shown by real complex networks.

**Shortest path length:** the shortest path length, or distance, $\ell_{ij}$ between vertices $i$ and $j$ is the length (in number of edges) of the shortest path joining $i$ and $j$. The shortest path length thus represents a measure of the distance between pairs of vertices. The average shortest path length $\langle \ell \rangle$ is the average of the shortest path length over all pair of vertices in the network, i.e.,

$$\langle \ell \rangle = \frac{2}{N(N-1)} \sum_{i<j} \ell_{ij},$$

where $N$ in the total number of vertices in the network.

**Small-world property:** property shown by many real complex networks which exhibit a small value of the average shortest path length $\langle \ell \rangle$, increasing with network size logarithmically or slower. This property is in stark contrast with the larger diameter of regular lattices, which grows algebraically with lattice size.

**Transitivity of a network**: propensity of two nodes in a network to be connected by an edge if they share a common neighbor.

**Strength of a node**: the sum of weights of the edges incident on a vertex.

**Tree**: a network that has as many edges as vertices minus one and is connected, i.e., a walk from one node can reach any other node in the network.

**Weighted network**: a network whose links are characterized by different capacities, or weights, defining the strength of the interaction between the nodes they connect.

**Word association network:** a network where vertices are words and a link connects a cue word with the word that is produced as response.

**BOX I: NETWORK MODELS**

The Erdös-Rényi random graph model (Fig. 2 and 3) has been the paradigm of network generation for a long time. It considers $N$ isolated nodes connected at random, in which every link is established with an independent connection probability $p$ [150]. The result is a graph with a binomial degree distribution, centered at the average degree, and little clustering. The availability of large-scale network data made clear that different models were needed to explain the newly observed properties, in particular a large clustering coefficient and a power-law distributed degree distribution [8]. The Watts-Strogatz model is one attempt to reconcile the high clustering characteristic of ordered lattices and small shortest paths lengths observed in complex networks [20]. In this model, in an initially ordered lattice, some edges are randomly rewired. For a small rewiring probability, clustering is preserved, while the introduction of a few shortcuts greatly reduces the network diameter. The Barabási-Albert model (Fig. 2) represents a first explanation of the power-law degree distributions found in many complex networks (Fig. 3) [21]. It is based on the principle of growth and preferential attachment: at each time step a new node enters the network and connects to old nodes proportionally to their degree; therefore 'richer nodes' (nodes with higher degree) 'get richer'. This rule leads to a degree distribution scaling as $P(k) \sim k^{-3}$. Exponents different from $3$ can be found, for example, by allowing for edge rewiring [22]. Other growth models displaying power-law degree distributions have been considered, involving mechanisms such as duplicating a node and its connections, with some edge rewiring [79, 151] or random growth by adding triangles to randomly chosen edges [152]. Non-growing alternatives to the origin of a scale-free topology have applied optimization mechanisms, seeking an explanation in terms of trade-offs, optimizing the conflicting objectives pursued in the set up of the network. Such models, elaborating on the highly optimized tolerance framework [153], find examples in the class of heuristically optimized trade-off (HOT) network models [154]. Other approaches, such as the class of models with 'hidden variables' [155] represent a generalization of the classical random graph in which the connection probability depends on some non-topological (hidden) variable attached to each edge. The proper combination of connection probability and hidden variables distribution can lead to a scale-free topology, without reference either to growth or preferential attachment [156].

**BOX II: COMPUTING WITH NETWORKS**

The central tenet of cognitive science is that thought is computation; and hence that the enormously rich network of neurons that composes the human brain is a computational device. Thus, a central intellectual challenge for cognitive science is to understand how networks of simple neurons-like units can carry out the spectacularly rich range of computations that underlie human thought, language, and behavior. *Connectionism*, or *parallel distributed processing* (see [157] for the historical pedigree) use networks composed of simplified neural processing units, where adjustments of the connections between units allow the models to learn from experience. This approach has been applied to many aspects of cognition from cognitive development [158] to language [159], including connectionist implementations [160] of symbolic semantic networks [115]. In parallel, an active tradition has aimed to provide computational models of actual neural circuitry; such models are more biologically realistic, but typically focus less on abstract cognitive tasks, and more on elementary processes of learning, early visual processing, and motor control [161].

Since the 1980s, there has been increasing interest in related, but distinct, research program, on using networks to *represent*, make *inferences* over, and *learn*, complex probability distributions [162]. In such probabilistic graphical models, nodes correspond to elementary states of affairs; and links encode probabilistic relationships, and even causal connections [163], between states of affairs. These models have proved to be powerful tools for artificial intelligence and machine learning, as well as the basis for many models in Bayesian cognitive science (e.g., [164]). Crucially, inference and learning in such models typically requires no "supervision"---nodes modify their level of activity in response to activity on incoming links; the strength of a link is modified in response to signals at the nodes that it connects.

In both connectionist networks and probabilistic graphical models, the network itself autonomously carries out inference and learning. However, the possible relationship between biological neural networks and these classes of psychological network models is less well understood. One suggestion is that neuromodulation, such as long-term potentiation (activity-dependent synaptic strengthening) corresponds to strengthening a 'connection' in a computational network; and more concretely the detection of "prediction error" (crucial in many network learning models) relates to activity of the dopamine system [165]; moreover, populations of neurons, and network operations over these, may implement probabilistic calculations (e.g., [166]). Nonetheless, understanding how networks can *compute* remains a central challenge for the cognitive and brain sciences.

**BOX III: DYNAMICAL PROCESSES ON NETWORKS**

Processes taking place upon networks are widespread across a large number of domains, from epidemics spreading through the airplane transportation network, to gossip spreading through networks of acquaintances [102]. In all cases, the topological properties of the underlying networks play a crucial role in the behavior of the process, and extremely simple models can provide vital insights into large classes of apparently distant phenomena. This is why the study of processes occurring on network has recently garnered a lot of attention also in cognitive science. In the main text, we have seen how the structure of the social network affects the spreading of a linguistic innovation [15], while random walk processes have been used in different contexts, from word association experiments [122] to language modeling [116].

The random walk is an ideal example to understand the insights that studying an apparently trivial process can provide. At each time step, a particle (the walker) hops from the node it occupies to a randomly selected neighboring node. The properties of such simple dynamics are enlightening in many respects. For example, it turns out that the so-called occupation probability $\rho_i$ of the walker, i.e. the asymptotic probability to find it on node $i$, is simply proportional to the degree $k_i$ of that node, i.e., $\rho_i \sim k_i$ in a connected network [167]. This node degree also turns out to be crucial in many more complex situations [8]. Other important properties, particularly relevant for the issues of searching and spreading in networks, are *mean first-passage time* (MFPT) and *coverage* [168]:

- The MFPT $\tau_i$ of a node $i$ is the average time taken by the random walker to arrive for the first time at vertex $i$, starting from a random source. This corresponds to the average number of messages that have to be exchanged among the nodes to identify the location of vertex $i$. Interestingly, in typical cases, this time is proportional to the inverse of the occupation probability.
- The coverage $C(t)$ is defined as the number of different vertices that have been visited by the walker at time t, averaged for different random walks starting from different sources. The coverage can thus be interpreted as the searching efficiency of the network, measuring the number of different individuals that can be reached from an arbitrary origin in a given number of time steps.

## BOX IV: THE MININIMUM LINEAR ARRANGEMENT

The minimum linear arrangement problem consists in finding a sequential ordering of the vertices of a network that minimizes the sum of edge lengths [73]. If $\pi(v)$ is the position of vertex $v$ and an $u \sim v$ indicates that that vertices $u$ and $v$ are connected, the length of the edge $u \sim v$ is the absolute value of the difference of their positions, i.e. $|\pi(v) - \pi(u)|$. The sum of edge lengths is

$$D = \sum_{u \sim v} |\pi(v) - \pi(u)|.$$

In a tree of $n$ vertices, the mean distance between edges is $<d>=D/(2(n-1))$. Imagine that a tree has only three vertices that are labeled with numbers 1,2 and 3. Then there are only 3! = 6 possible linear arrangements of the vertices (Fig. 4 (a)) but the minimum $<d>$ (or equivalently the minimum $D$) is achieved by only two orderings, (1,2,3) and its reverse (3,2,1) with $<d>=1$ (Fig. 4 (a)). We say that these two orderings are minimum linear arrangements. $<d> = 1.5$ for the remainder of orderings.

In a star tree, where all vertices have degree one except one, i.e. the hub (Fig. 4 (b)), $D$ is determined by the position of the hub in the sequence. For that tree, the optimal placement of the hub is at center of the sequence [78].

The ordering of the words in the sentence of Fig. 4 (c), which yields $<d>=11/8 \approx 1.375$, is also a minimum linear arrangement, i.e. none of the 9! = 362880 permutation of the words of the sentences is able to achieve a smaller $<d>$ given the syntactic dependency tree of the sentence. Finding the minimum linear arrangement problem of a network is very hard computational problem [73] but if the network is a tree (e.g., Fig. 4 (c)), computationally efficient solutions exist [169, 170].

$<d>$ would grow linearly  ($<d>=(n+1)/3$) with the number of vertices, if vertices were ordered at random[72]. In contrast, $<d>$ grows sublinearly as a function of the number of vertices in real syntactic dependency trees [72].

$\langle k^2 \rangle$, the degree 2nd moment. determines the minimum value that $<d>$ could achieve, i.e. [78]

$$\langle d \rangle \geq \frac{n \langle k^2 \rangle}{8(n-1)} + \frac{1}{2}.$$

The worst case is a star tree (Fig. 4 (b)) with the maximum $\langle k^2 \rangle$ [78]. Therefore, the tendency to have "hubs" " (i.e. a high degree variance in degrees of different vertices) and a low $<d>$ are incompatible.

**BOX V: FRONTIERS IN NETWORK SCIENCE**

In the main text we have reviewed key contributions of network theory to cognitive science, highlighting that, along with the traditional study of properties of fixed networks (Section II), a recent wave considers also dynamical process upon networks (Section III). Here, we sketch a brief overview of some topics at the frontiers of network science [171], which may have a substantial impact on cognitive science and many other disciplines in the near future.

A first challenge concerns the problem of timescale separation. Traditionally, two limits have been considered in the study of dynamical processes on networks: either the network is considered to be effectively static, meaning that it evolves on a timescale much slower than the one of the process under consideration, or, on the contrary, it is described as rapidly varying with a pace that allows the process to perceive only the statistical properties of the graph, e.g., the degree distribution only. [8]. The issue is now to develop tools to describe what happens in the intermediate situations, i.e., when the timescale of the dynamical process is comparable to the rate of network evolution [172]. Real-world examples of this can be found in social and cognitive processes taking place on face-to-face interaction networks [98], or on online messenger sites such as Twitter [91].

The second challenge is deeply connected to the first, and goes one step further. What happens when the dynamical process co-evolves with the underlying network, so that both dynamics interact with each other through feedback mechanisms? Recent research has shown that, when this is the case, very interesting self-organization phenomena may arise, such as the possible fragmentation of social networks when links can be rewired depending on the dynamical state (i.e., the opinion) of the nodes (i.e., the individuals) they connect [173, 174].

Finally, apart from the challenges of describing, modeling and understanding complex networks, a further question is how they can be controlled [175]. Control theory offers important mathematical tools to address this question, but the network heterogeneity introduces nontrivial issues that have just started to be taken into account. Identifying driver nodes that can guide the system's entire dynamics over time, for example, might help engineering an observed system to perform desired function, or prevent malfunctioning. Interestingly, it turns out that such nodes tend not to be the hubs of the network [175].

**OUTSTANDING QUESTIONS:**

- Is network theory a framework that can unify the representation of structure across levels and domains in cognitive science and neighboring disciplines (e.g., from neural organization to knowledge representation)?

- To what extent do the underlying brain networks determine the properties of cognitive networks and vice versa? Which well-known properties of brain networks are also found at higher levels in cognitive networks and vice versa?

- What are their optimal values of path lengths and/or clustering for proper brain functioning, cognitive processing, or social dynamics? Do these optimal values depend on the cognitive domain? Do very low or high values of indicate pathology? If so, do such indicators apply across different explanatory levels: e.g., do the aberrant statistical properties of brain networks observed in Alzheimer's disease, schizophrenia or autism arise also at the cognitive level?

- Are the properties of the network structure in social interactions a key factor for the emergence of complex individual abilities such as language (e.g., syntax)? And conversely, to what extent are the properties of these social interactions determined by individual cognitive abilities (e.g., Dunbar's number)?

**TABLE I**

**Cognitive science through the eyes of network theory: Translation of cognitive science terms into network theory concepts.**

| COGNITIVE SCIENCE AND NEIGHBOURING FIELDS | NETWORK THEORY |
|---|---|
| Semantic field | Community in a network (e.g., word association network) [57, 58] |
| Island | Connected component [81] |
| Brain module | Community in a brain network [35] |
| Semantic memory | Semantic network [58] |
| Mental exploration (mental navigation without a target) | Random walk in a cognitive network [58, 116] |
| Tagging activity by users | Random walk in a mental semantic network [122] |
| Landmark (in a wayfinding problem) | Node with high closeness centrality [48] |
| Pathological brain or pathological cognition | Anomalous network metrics, e.g., clustering and path lengths [28] [46, 53] |
| Unfounded scientific authority, first-mover advantage | Rich-get-richer phenomenon on a citation network [87, 88]. |